\documentclass[11pt,twoside]{article}

\usepackage{asp2014}

\aspSuppressVolSlug
\resetcounters

\begin{document}

\title{A practical preconditioner for wide-field continuum imaging of radio interferometric data}

\author{Hertzog~L.~Bester,$^{1,2}$ Audrey~Repetti,$^3$ Simon~Perkins,$^{1,2}$ Oleg~M.~Smirnov,$^{2,1}$ and Jonathan~S.~Kenyon$^{2,1}$}
\affil{$^1$South African Radio Astronomy Observatory, Cape Town, Western Cape, South Africa; \email{lbester@ska.ac.za}}
\affil{$^2$Rhodes University, Makhanda (Grahamstown), Eastern Cape, South Africa}
\affil{$^3$Institute of Sensors, Signals and Systems, Heriot-Watt University, Edinburgh, United Kingdom}

\paperauthor{Hertzog~L~Bester}{lbester@ska.ac.za}{0000-0002-7348-2229}{South African Radio Astronomy Observatory}{Radio Astronomy Research Group}{Cape Town}{Western Cape}{7925}{South Africa}
\paperauthor{Audrey~Repetti}{a.repetti@hw.ac.uk }{}{Author2 Institution}{Author2 Department}{City}{State/Province}{Postal Code}{Country}
\paperauthor{Simon~Perkins}{simon.perkins@gmail.com}{}{South African Radio Astronomy Observatory}{Radio Astronomy Research Group}{Cape Town}{Western Cape}{7925}{South Africa}
\paperauthor{Oleg~M~Smirnov}{osmirnov@gmail.com}{}{Rhodes University}{Radio Astronomy Techniques and Technologies}{Makhanda (Grahamstown)}{Eastern Cape}{6140}{South Africa}
\paperauthor{Jonathan~S~Kenyon}{osmirnov@gmail.com}{}{Rhodes University}{Radio Astronomy Techniques and Technologies}{Makhanda (Grahamstown)}{Eastern Cape}{6140}{South Africa}



\begin{abstract}
The celebrated CLEAN algorithm has been the cornerstone of deconvolution algorithms in radio interferometry almost since its conception in the 1970s. For all its faults, CLEAN is remarkably fast, robust to calibration artefacts and in its ability to model point sources. We demonstrate how the same assumptions that afford CLEAN its speed can be used to accelerate more sophisticated deconvolution algorithms. 
\end{abstract}





\section{Introduction}
There is by now a wealth of research dedicated to improving interferometric imaging techniques (eg. \citet{thouvenin2020parallel,arras2020comparison}). Despite the fact that many algorithms are capable of superior imaging performance compared to CLEAN, they have not been widely adopted by the radio astronomy community. One of the main reasons for this is increased computational complexity. In particular, compared to CLEAN, most competitor algorithms require many more applications of the full measurement operator which is often the most expensive aspect of the imaging problem. 

In what follows we demonstrate how, by approximating the Hessian of the data fidelity term (i.e. negative log-likelihood) as a convolution with the point spread function, it is possible to develop an effective preconditioner for a proximal gradient based imaging algorithm. The resulting algorithm, dubbed preconditioned forward-backward clean (\emph{pfb-clean}), often requires very few applications of the full measurement operator and is therefore particularly suited to imaging radio interferometric data in the regime where the data size is much larger than that of the image. 

\section{Methodology}
The interferometric imaging problem amounts to finding an estimate of an unknown discretised image of the sky $x$ from incomplete measurements $V$. The corresponding ill-posed inverse problem is given by
\begin{equation}
    V = R x + \epsilon, ~~ \epsilon \sim \mathcal{N}\left(0, \Sigma\right),
\end{equation}
where $R$ is the linear measurement operator and $\epsilon$ is a realisation of Gaussian noise with (assumed diagonal) covariance matrix $\Sigma$. During continuum imaging, it is typical to assume that flux varies slowly as a function of frequency and that it is sufficient to reconstruct the image at a much lower frequency resolution than that of the data. Thus (for an ideal unpolarised interferometer) $R$ is a map $ R: \mathbb{R}^{n_b \times n_p} \rightarrow \mathbb{C}^{n_\nu \times n_d} $ with $ n_b \ll n_\nu$ where $n_\nu$ is the number of frequency channels, $n_d$ the number of rows per channel, $n_b$ the number of imaging bands and $n_p$ the number of image pixels per band. This map can be implemented efficiently as a 2D non-uniform fast Fourier transform per imaging band with a correction that accounts for wide field effects (see eg. \citet{arras2020efficient}). We use the dask wrappers of the \verb!wgridder! \citet{arras2020efficient} in \verb!codex-africanus! (\href{https://github.com/ska-sa/codex-africanus}{https://github.com/ska-sa/codex-africanus}) to implement the measurement operator throughout and use \verb!dask-ms! (\href{https://github.com/ska-sa/dask-ms}{https://github.com/ska-sa/dask-ms}) as the data access layer (see \citet{O8-131_adassxxx}). 

Since the noise is Gaussian, the data fidelity term and its gradient are given by
\begin{eqnarray}
    f(x) &=& \frac12(V - Rx)^\dagger \Sigma^{-1} (V-Rx) \\
    \nabla_x f &=& -R^\dagger \Sigma^{-1}(V - Rx) = R^\dagger \Sigma^{-1} R x - I^d,
    \label{exactgrad}
\end{eqnarray}
where the dirty image $I^d = R^\dagger \Sigma^{-1} V$ is the data projected into image space which, since $n_b \times n_p \ll  n_\nu \times n_d$, results in a significant dimensionality reduction. Unfortunately, the presence of the Hessian $R^\dagger \Sigma^{-1} R$ means that gradient computations are inevitably expensive. However, by using the Fourier convolution theorem and the approximate Fourier nature of $R$, we can approximate the gradient as
\begin{equation}
    \nabla_x f \approx I^{psf} \star  x ~ - ~ I^d,
    \label{approx_grad}
\end{equation}
where $I^{psf} \star$ denotes convolution with the point spread function (PSF) of the instrument. It is not, in general, sufficient to use \eqref{approx_grad} to approximate the gradient because it neglects wide-field effects when the array is not coplanar.

The interferometric imaging problem amounts to solving problems of the form
\begin{equation}
    \underset{x}{\mbox{minimise}} \quad f(x) + r(x),
    \label{objective}
\end{equation}
where $r(x)$ is some regularising function. One popular choice for solving \eqref{objective} when $r(x)$ is not smooth is the forward-backward proximal gradient algorithm. Such algorithms usually require many gradient evaluations. Thus, we propose using a preconditioned variant of the forward-backward algorithm (see \citet{repetti2020variable}) that can drastically reduce the required number of gradient evaluations. The algorithm alternates between gradient (forward) and backward steps given, respectively, by
\begin{eqnarray}
    x_{k+\frac12} &=& x_k - \gamma U^{-1} \nabla_x f(x_k), \label{pfbupdate} \\
    x_{k+1} &=& \mbox{prox}^U_{\gamma r}(x_{k+\frac12}) = \underset{x}{\mbox{argmin}} ~~ r(x) + \frac{1}{2\gamma}(x-x_{k+\frac12})^\dagger U(x-x_{k+\frac12}).
    \label{pfbproject}
\end{eqnarray}
The choice of $U$ is not arbitrary but subject to some mild constraints (see \citet{repetti2020variable} for full details). The similarity of \eqref{pfbupdate} to the standard Gauss-Newton update rule suggests using the Hessian of $f(\cdot)$ as the preconditioner. Intuitively, since the inverse Hessian approximates the covariance in the update, the requirement in \eqref{pfbproject} would then simply mean that we have to keep track of this approximate covariance during the backward step. However, simply using $U = R^\dagger \Sigma^{-1} R$ is i) not possible because it is not invertible and ii) not practical because it would require too many applications of the measurement operator. Fortunately, a valid preconditioner can be obtained by adding a small multiple of the identity to the approximate Hessian i.e.
\begin{equation}
    U = I^{psf} \star ~ + ~ \sigma \mbox{Id},
\end{equation}
where $\mbox{Id}$ is the identity and $\sigma > 0$ is used to regularise the inversion of $U$. We can then exploit efficient matrix vector products of the operator $U$ and use the conjugate gradient algorithm to compute \eqref{pfbupdate}. Similarly, the primal dual algorithm of \citet{condat2012} can be used to solve \eqref{pfbproject} for a large class of regularisers. Importantly, the preconditioning strategy makes it possible to use large step sizes (typically $\gamma \lesssim 1$) so that very few exact gradient evaluations are required.

\articlefigure[width=.9\textwidth]{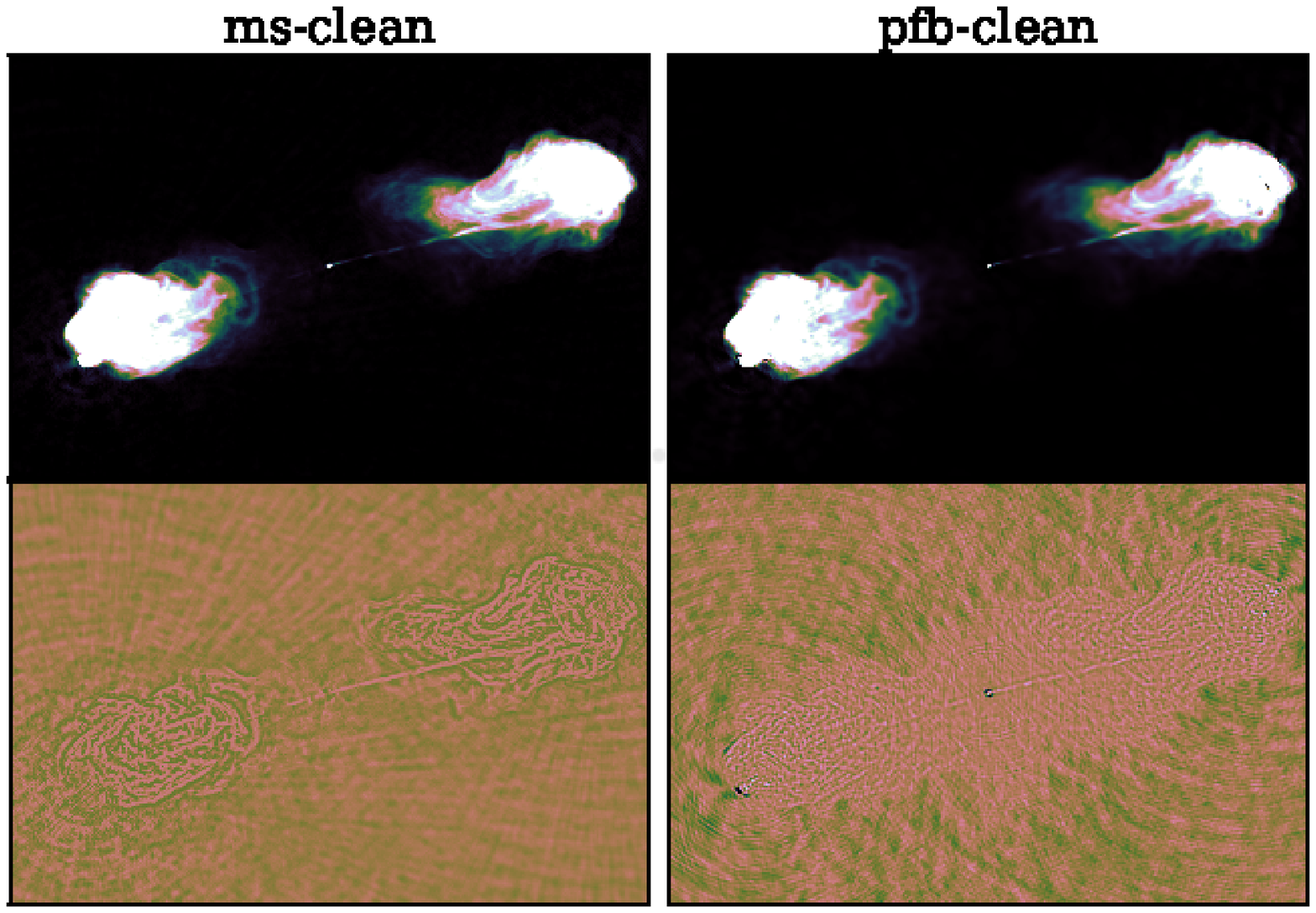}{P5-58_f1}{Comparison of Cygnus A reconstructed with wsclean (left) and pfb-clean (right). The top panel shows the science data products while the bottom panel shows the corresponding residual images.}

\section{Results and discussion} 
Figure~\ref{P5-58_f1} shows a Stokes I multi-frequency synthesis image of Cygnus A reconstructed from S band VLA data in the A, B, C and D configurations \citet{sebokolodi2020wideband} using a regulariser of the form
\begin{equation}
    r(x) = \lambda \| \Psi^\dagger x \|_{2,1} + \iota_{[0,+\infty]}(x),
    \label{prior}
\end{equation}
where $\| \cdot \|_{2,1}$ is the $l_{2,1}$ norm,  $\iota_{[0,+\infty]}(\cdot)$ is an indicator function used to enforce positivity of the image, $\lambda$ sets the strength of the $l_{2,1}$ prior and $\Psi$ is a dictionary containing the Dirac and first six Daubechies wavelets. We compare our result to that obtained with the multi-scale CLEAN (ms-clean) algorithm (with default multi-scale settings and using auto-masking) in \verb!wsclean! \citet{Offringa_2017}. For both applications, we create $1728\times 1200$ pixel images with a cell size of 0.125 arc-seconds in 8 imaging bands, one for each 128 channel spectral window. This results in oversampling the Fourier grid at approximately twice the Nyquist rate at the highest frequency. We observed that Briggs weighting with a robustness factor of about -1 is required to approximately match the resolution obtained with pfb-clean using natural weighting. However, the resolution of pfb-clean is limited in this particular example because of the presence of calibration artefacts. In general, since the $\Psi$ dictionary is more expressive than the one utilised by ms-clean, pfb-clean is more sensitive to calibration artefacts. These can be partially mitigated by dialling up the value of $\lambda$ but this comes at the price of not being able to capture some of the finer morphological features. Also note that, although pfb-clean produces worse residuals, these are naturally weighted residuals corresponding to the actual science data product which respects the positivity of flux, unlike the residuals produced by ms-clean. Finally, in this example, ms-clean requires eight gradient evaluations to reach the final threshold of $500~ \mu \mbox{Jy/beam}$ and we use the same number of forward-backward steps for pfb-clean. The computational cost of pfb-clean is still significantly higher because of the sub-iterative nature of solving \eqref{pfbupdate} and \eqref{pfbproject}. However, the main computational cost during these steps are image sized FFT's and wavelet decompositions which can be efficiently parallelised. There is also the possibility of sub-dividing the image into facets to further reduce the computational cost.

\acknowledgements We thank Philipp Arras, Ming Jiang, Tim Molteno, Martin Reinecke and Yves Wiaux for useful discussions and Richard A. Perley for providing Cygnus A observations with the VLA. The research of OS is supported by the South African Research Chairs Initiative of the Department of Science and Technology and National Research Foundation. This work was supported in part by the Swiss-South Africa Joint Research Program (IZLSZ2170863/1).

\bibliography{P5-58}  

\begin{thebibliography}{}
\expandafter\ifx\csname natexlab\endcsname\relax\def\natexlab#1{#1}\fi
\expandafter\ifx\csname url\endcsname\relax
  \def\url#1{\texttt{#1}}\fi
\expandafter\ifx\csname urlprefix\endcsname\relax\def\urlprefix{URL }\fi
\providecommand{\eprint}[2][]{\url{#2}}

\bibitem[{Arras et~al.(2020{\natexlab{a}})Arras, Bester, Perley, Leike,
  Smirnov, Westermann, \& Enßlin}]{arras2020comparison}
Arras, P., Bester, H.~L., Perley, R.~A., Leike, R., Smirnov, O., Westermann,
  R., \& Enßlin, T.~A. 2020{\natexlab{a}}, Comparison of classical and
  bayesian imaging in radio interferometry. \eprint{2008.11435}

\bibitem[{Arras et~al.(2020{\natexlab{b}})Arras, Reinecke, Westermann, \&
  Enßlin}]{arras2020efficient}
Arras, P., Reinecke, M., Westermann, R., \& Enßlin, T.~A. 2020{\natexlab{b}},
  Efficient wide-field radio interferometry response. \eprint{2010.10122}

\bibitem[{Condat(2013)}]{condat2012}
Condat, L. 2013, {Journal of Optimization Theory and Applications}, online
  first, to appear.
  \urlprefix\url{https://hal.archives-ouvertes.fr/hal-00609728}

\bibitem[{Offringa \& Smirnov(2017)}]{Offringa_2017}
Offringa, A.~R., \& Smirnov, O. 2017, Monthly Notices of the Royal Astronomical
  Society, 471, 301–316.
  \urlprefix\url{http://dx.doi.org/10.1093/mnras/stx1547}

\bibitem[{{Perkins} et~al.(2021)}]{O8-131_adassxxx}
{Perkins}, S.~J., et~al. 2021, in ADASS XXX, edited by J.-E. {Ruiz}, \&
  F.~{Pierfederici} (San Francisco: ASP), vol. TBD of ASP Conf. Ser., 999 TBD

\bibitem[{Repetti \& Wiaux(2020)}]{repetti2020variable}
Repetti, A., \& Wiaux, Y. 2020, Variable metric forward-backward algorithm for
  composite minimization problems. \eprint{1907.11486}

\bibitem[{Sebokolodi et~al.(2020)Sebokolodi, Perley, Eilek, Carilli, Smirnov,
  Laing, Greisen, \& Wise}]{sebokolodi2020wideband}
Sebokolodi, M.~L., Perley, R., Eilek, J., Carilli, C., Smirnov, O., Laing, R.,
  Greisen, E., \& Wise, M. 2020, A wideband polarization study of cygnus a with
  the jvla. i: The observations and data. \eprint{2009.06554}

\bibitem[{Thouvenin et~al.(2020)Thouvenin, Abdulaziz, Jiang, Dabbech, Repetti,
  Jackson, Thiran, \& Wiaux}]{thouvenin2020parallel}
Thouvenin, P.-A., Abdulaziz, A., Jiang, M., Dabbech, A., Repetti, A., Jackson,
  A., Thiran, J.-P., \& Wiaux, Y. 2020, Parallel faceted imaging in radio
  interferometry via proximal splitting (faceted hypersara): when precision
  meets scalability. \eprint{2003.07358}

\end{thebibliography}


\end{document}